\documentclass[final]{aipproc}
\pdfoutput=1

\layoutstyle{6x9}
\usepackage{xspace} 
\usepackage{graphicx}

\graphicspath{{./figures/}}
\newcommand{\figurewidth}{3.0in}

\newcommand{\sstt}        {$\sin^2 2 \theta$\xspace}
\newcommand{\dms}         {$\Delta m^2$\xspace}

\newcommand{\numu}        {$\nu_{\mu}$\xspace}

\begin{document}

\title{Why understanding neutrino interactions is important for
  oscillation physics}


\author{Christopher W. Walter}
{address={Department of Physics, Duke University, Durham, NC 27708 USA}}

\begin{abstract}

  Uncertainties in knowledge of neutrino interactions directly impact
  the ability to measure the parameters of neutrino oscillation.
  Experiments which make use of differing technologies and neutrino
  beams are sensitive to different uncertainties.

\end{abstract}

\maketitle

\section{Introduction}

Strong evidence for neutrino oscillation and the existence of neutrino
mass exists from atmospheric neutrinos~\cite{Ashie:2005ik}, solar
neutrinos~\cite{Fukuda:2002pe,Ahmad:2002jz}, reactor
experiments~\cite{Araki:2004mb}, and long baseline oscillation
experiments~\cite{Ahn:2006zz}.  The recent results from the MiniBooNE
experiment~\cite{AguilarArevalo:2007it} have confirmed our standard
model of neutrino oscillations.  The picture of neutrino physics we
have extracted is that there are three active neutrinos with two mass
splittings.  Of the three mixing angles needed to mix the three mass
states together, two are large or near maximal and one is small and
possibly zero.

Theoretical attempts to explain why the neutrino masses are so small
and their mixings are large often rely on physics at the GUT scale
(for a recent discussion see ~\cite{unknown:2005br}).  One of the most
popular ideas, known as the See-Saw mechanism~\cite{seesaw}, coupled
with CP violation in neutrinos produces
leptogenesis~\cite{Fukugita:1986hr}, where a lepton matter/antimatter
asymmetry caused by the decay of heavy neutrinos is converted into a
baryon asymmetry and explains why today we live in a matter dominated
universe.  To explore these ideas there are a set of questions which
need to be experimentally addressed.  These are:

\begin{itemize}
\item What is the relative pattern of masses of the known neutrino
  mass differences?
\item What is the size of the one neutrino mixing angle that has not
  been measured?  Is it large enough to allow us to eventually measure
  the violation of CP if it exists?
\item Do neutrino violate CP symmetry?
\item Unlike quarks the neutrino mixing angles that have been measured
  are large, some possibly even maximal. Are the largest angles really
  maximal and what would that imply?
\end{itemize}

\noindent
Unfortunately, we do not live in a world where we can cleanly interact
neutrinos off of single quarks.  Quarks come bundled inside of nucleons
which themselves are found in the nucleus.  For this reason, in order
to extract the information about neutrino oscillations and masses we
wish from our experiments, we must also understand the physics of
neutrino interactions inside of nuclear material.

\section{Cross Sections}

Since the first NuInt01 meeting~\cite{Morfin:2002fk} we have made a
lot of progress in this field.  However, there are many outstanding
questions, some of them quite basic.  One example, which has been
noted experimentally by both the K2K and MiniBooNE experiments, is the
unexpected suppression at low $Q^2$ of charged-current quasi-elastic
interactions.  At this meeting we saw new work from the MiniBooNE
collaboration to address this issue~[See these proceedings].

\begin{figure}[!htb]
  \includegraphics[width=4.5in]{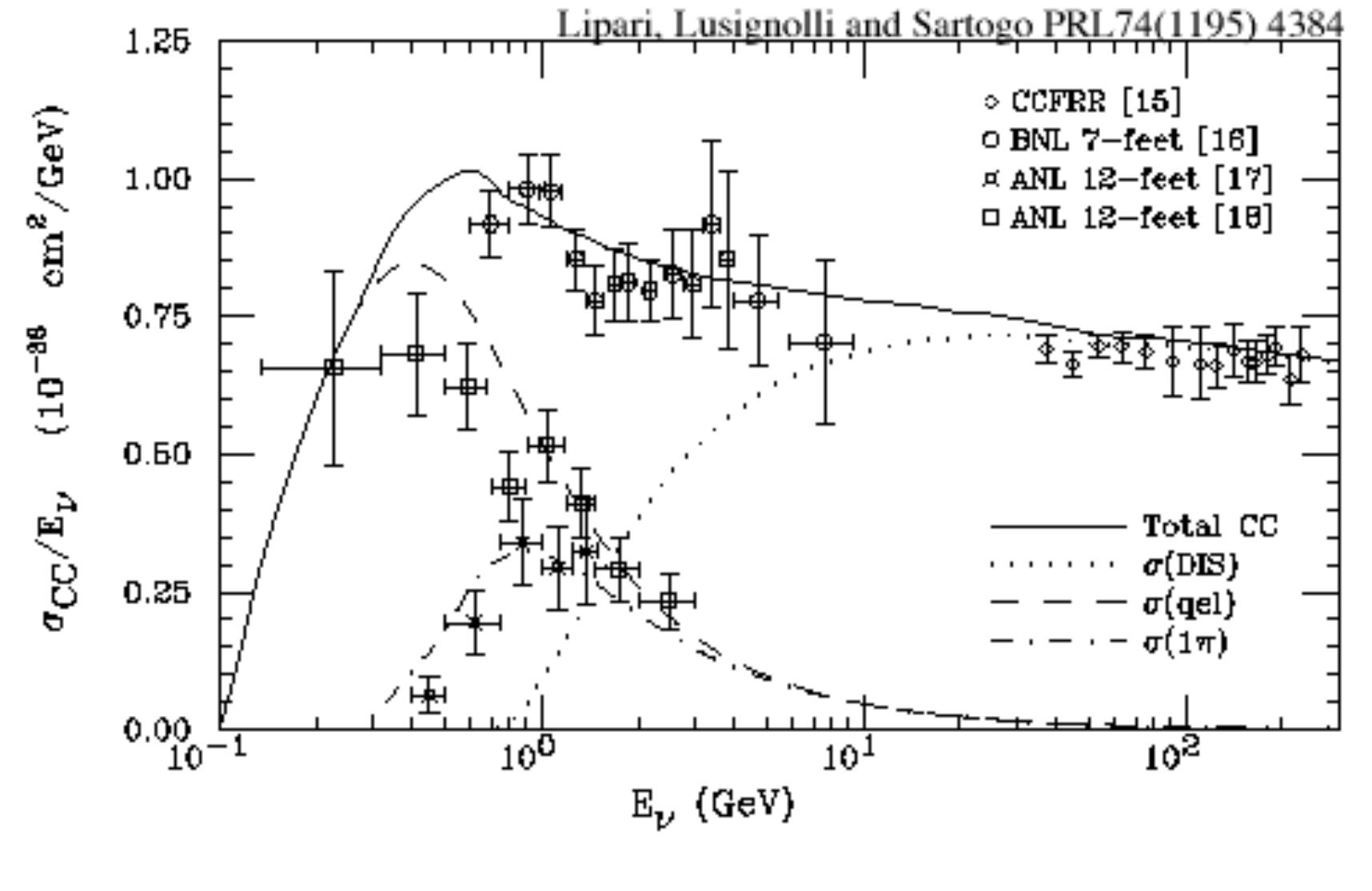}
  \caption{The neutrino nucleon cross section as a function of
    neutrino energy.  Below 1~GeV the cross section is dominated by
    quasi-elastic interactions.  For these interactions the neutrino
    energy can be reconstructed using only the outgoing lepton.}
  \label{fig:cross-section}
\end{figure}

There are also subtler effects that take place in the nucleus, some
dependent on the type of nucleus the interaction takes place in.  The
neutrino interaction cross-sections are shown (along with some data)
in figure~\ref{fig:cross-section} which is taken
from~\cite{Lipari:1994pz}.

Different experiments sample different parts of this figure.  For
example, the T2K experiment~\cite{Itow:2001ee} has a beam which is
peaked below 1~GeV and is therefore dominated by quasi-elastic
interactions.  The NoVa~\cite{Ambats:2004js} experiment on the other
hand uses neutrinos in the few-GeV and above range.

An important way to mitigate the problems due to uncertainties in
these cross-sections is to use both a near and far detector to measure
the interactions before and after interactions.  However, what is
measured in each detector is the flux$\times$cross-section and any
differences in detector efficiency, flux, or cross-section
between the two detectors will be convolved with the errors due to the
nuclear effects and will not completely cancel.

The are several types of interaction uncertainties to consider, and
which ones are important depend both on the detector technology being
used, and the physics analysis being performed.  Future experiments
which wish to probe CP violation will also make use of anti-neutrino
beams, and so we must understand the cross-sections of those
anti-neutrinos on nuclear material as well. The first results from
high statistics anti-neutrino running were shown by the MiniBooNE
collaboration in this meeting~[See these proceedings].

\section{Reconstructing Neutrino Events in Detectors}
\label{sec:detectors}

Neutrino oscillation analyses can be broadly separated into two
classes: searches for neutrino disappearance and appearance.  In
disappearance experiments, a neutrino flavor oscillates into another
neutrino flavor for which there is not enough energy for a
charged-current interaction to take place and produce a lepton. The
measured effect is a distortion in the observed energy spectrum at the
far detector.  In an appearance experiment, one searches for the
appearance of a flavor at the far detector which was not present in
the initial beam.  Different experiments use different techniques
depending both on the detector technology used and the energy of the
incoming neutrinos.  Here, I touch on three illustrative examples
which show uncertainties in neutrino interactions can affect
oscillation results.

\begin{enumerate}
\item Reconstruction of the neutrino energy spectrum in large
  Water Cherenkov detectors.
\item Reconstruction of the neutrino energy spectrum in large
  calorimetric detectors.
\item Identification and reconstruction of tau neutrino events in
  large hybrid tracking/emulsion detectors.
\end{enumerate}

\subsubsection{Water Cherenkov detectors: the effect of non-QE interactions}
\label{sec:water}
\label{sec:nonqe-effect}

Water Cherenkov detectors like Super-Kamiokande~\cite{Fukuda:2002uc}
achieve a large mass by using water both as a target and active
detector element.  However, because of the nature of the Cherenkov
process not all particles produced in neutrino interactions are
visible in a water Cherenkov detector.  Fortunately, if the reaction
is quasi-elastic(QE) the kinematics of the event and the incoming
neutrino energy can be reconstructed using only the energy and angle
with respect to the beam of the produced lepton.
Equation~\ref{eq:recon-energy} shows the relationship between the
incoming neutrino energy and the reconstructed momentum of the
produced lepton.

\begin{equation}
  E_\nu = { m_N E_\mu -  m_u^{2}/2
    \over  m_N - E_\mu + p_\mu \cos(\theta_\mu) } ,
  \label{eq:recon-energy}
\end{equation}

\begin{figure}[!htb]
  \includegraphics[width=\figurewidth]{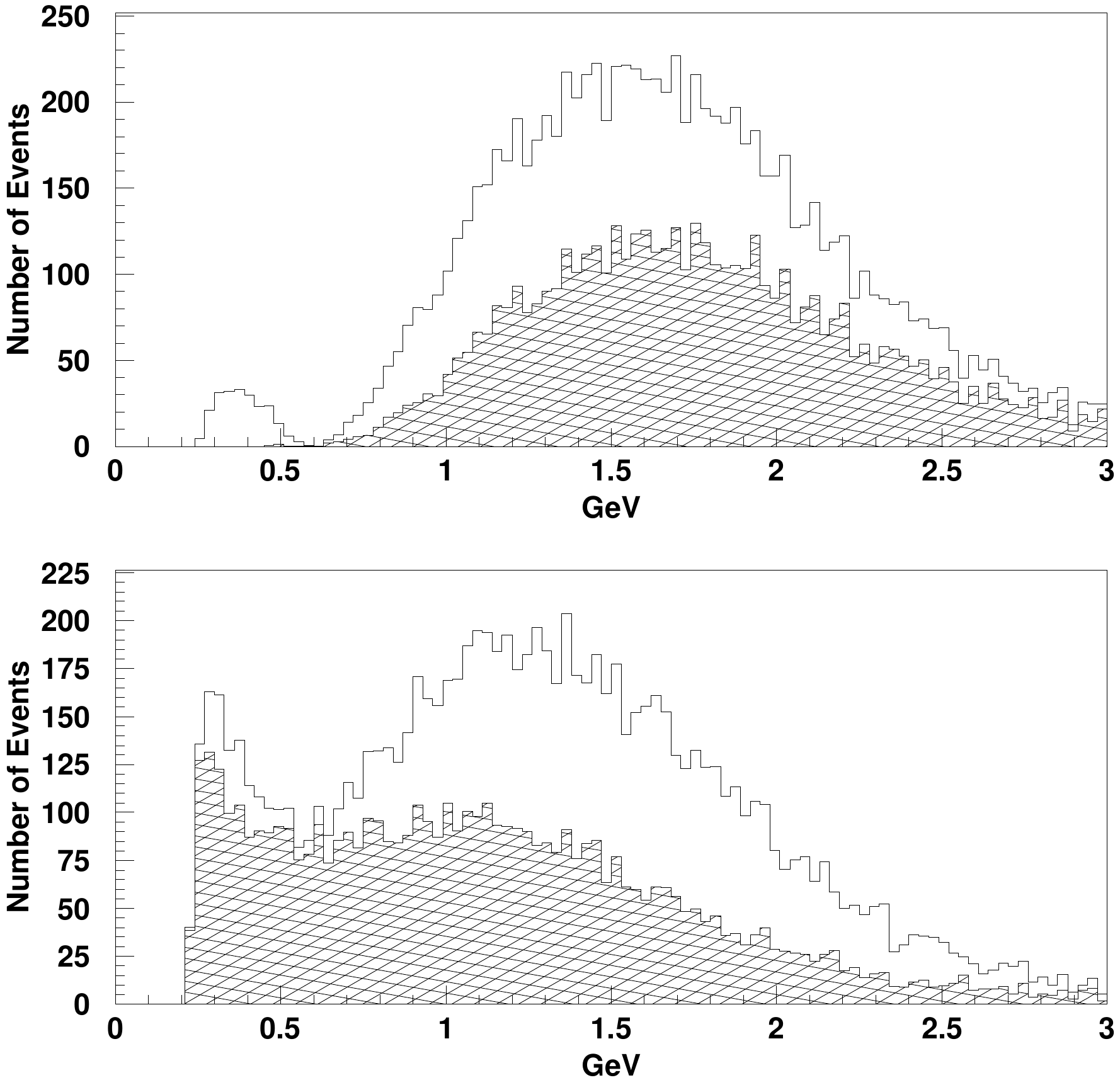}
  \caption{The top panel shows the Monte Carlo K2K spectrum at Super-K
    with oscillations applied.  The oscillation dip at 700~MeV
    maximally suppresses the flux of \numu neutrinos.  The non-QE
    interactions(hatched region) are un-effected by oscillations
    because their energy is too high. The bottom panel shows the same
    thing using reconstructed energy.  The non-QE interactions ``fill
    in'' the oscillation dip.}
  \label{fig:oscillation}
\end{figure}

\noindent
Unfortunately, those events which are not due to quasi-elastic
interactions will have their energies systematically underestimated.
Figure~\ref{fig:oscillation} shows the effect on mis-reconstruction on
an oscillation experiment.

For this reason, it is quite important to accurately model the
fraction and shape of this non-QE ``background''.  The parameter \sstt
determines the overall normalization of the oscillation suppression,
with \sstt=1 resulting in a complete suppression of the flux.  If the
amount of non-QE interactions is not-properly modeled, then the
overall suppression in the oscillation region will not be modeled
properly either, and the less than maximal suppression will be
incorrectly interpreted as a \sstt less than unity.  

\subsubsection{Calorimetric detectors: the effect of pion absorption}

Large calorimetric detectors like MINOS~\cite{Michael:2006rx} use a
different reconstruction technique and are most sensitive to a
different set of interaction uncertainties.  One advantage of a
colorimetric detectors relative to water Cherenkov detectors is that
all of the particles are in principle visible.  However, in order to
range out the particles in high energy interactions heavy materials
such as steel are often used.  In the MINOS experiment a large
fraction of the events come from deep inelastic scattering and in
order to reconstruct the neutrino energy the energy of the outgoing
lepton plus all pions and secondary particles in the shower must be
added up.  Equation~\ref{eq:recon-energy-calor} shows the relationship
between the incoming neutrino energy and the reconstructed momentum of
the produced lepton and energy of any associated shower,

\begin{equation}
  E_\nu = E_\mu + E_{shower}.
  \label{eq:recon-energy-calor}
\end{equation}

This use of this technique means that any unaccounted for loss in
energy of the shower will directly translate into an error on the
reconstructed energy scale. The can happen as pion are absorbed in the
steel planes and within the iron nuclei themselves.  This energy scale
uncertainty caused by hadronic interactions is currently on the order
of 10\% in the MINOS experiment and is the second largest systematic
error on the measured \dms.  The effect on internuclear interactions in
the MINOS experiment was nicely demonstrated my M. Kordoski at the
NuInt05 meeting~\cite{Cavanna:2006ry}.

\subsubsection{Tracking detectors: The effect of charm production}

Hybrid emulsion tracking detectors like the OPERA
experiment~\cite{Marteau:2007uf} face a very different set of problems
and challenges. OPERA is an appearance experiment and is looking for
the tell-tale kink of a tau decay in their emulsion.  Tracking
chambers are used to guide an automatic emulsion scanning system back
to the vertex of the event. At this point a kinematic reconstruction
and topological analysis is done to attempt to identify the small
number of tau events expected in the sample.

The main backgrounds for this sort analysis include hadronic
re-interactions which can cause kinks in tracks that look like decays
and charm decays which can be misidentified as having tau-decay
topology.  Future decreases in the uncertainties in charm production
cross-sections would decrease the uncertainty on this background.

\section{Modeling Interactions in the Nucleus}

All of the effects motioned above must be modeled in our neutrino
interaction Monte Carlos.  The previous examples were only a few of
the effects that must be considered. Intense theoretical and modeling
work is addressing a whole suite of issues in neutrino interaction
physics.  Many of these issues are addressed more fully in this
volume.  Due to lack of space, I only list many of the more relevant
issues here:

\begin{itemize}
\item The modeling of the quasi-elastic cross-section and axial mass.
\item New work on non-dipole nuclear form factors.
\item Models of resonant and coherent pion production.
\item Deep inelastic scattering and the transition to the resonance region.
\item Proper modeling of final states due to the Pauli exclusion principle.
\item The use of spectral functions to model binding energy and lepton
  momentum.
\item The re-scattering of final state particles in the nucleus.
\item The modification of parton distribution functions in the
  presence of other nucleons.
\end{itemize}

One item in this list above deserves special mention since it was the
subject of intense discussion in this workshop.  The K2K experiment
has measured a striking deficit in the amount of charge-current
coherent pion production~\cite{Hasegawa:2005td}.  The amount of
neutral current production on the other hand seems to agree with the
theoretical models.  The amount of pion production in neutrino beams is
of importance to the next generation of long-baseline experiments
since mis-identified neutral pions are an important source of background
in the search for electron neutrino appearance.

New theoretical work presented at this workshop can explain at least
some of this deficit in the charged current channel by correctly
incorporating the mass of the final state lepton in the calculations.

\section{Conclusions}

Uncertainties in neutrino interactions are a important source of
systematic errors when trying to make precision measurements of
neutrino oscillation parameters.  Not only must the effects themselves
be understood and properly modeled but the uncertainties on these
effects need to be properly accounted for in analyses.

The NuInt series has been extremely important both in addressing these
issues and in fostering new experimental collaborations.  The new data
we see in this meeting and we soon expect to see from dedicated
interaction experiments will be a crucial piece of the world-wide
effort to untangle the unknown physics of neutrino oscillations.

\bibliographystyle{aipproc}   
\bibliography{references}

\end{document}